\documentclass[twocolumn]{aa}  
\usepackage{ mathrsfs }
\usepackage{graphicx}
\usepackage{txfonts}
\usepackage{amssymb}
\usepackage{natbib}
\usepackage{supertabular}
\usepackage{longtable}
\usepackage{graphicx}
\usepackage{lscape}
\usepackage{ulem}
\usepackage{multirow}
\usepackage{rotating}
\usepackage{dcolumn}
\usepackage[breaklinks,colorlinks,linkcolor=red,citecolor=blue,urlcolor=blue]{hyperref}
\newcolumntype{d}{D{.}{\cdot}{-1}}

\begin{document}

   \title{The spectroscopic indistinguishability of red giant branch and red clump stars}

%   \subtitle{I. Overviewing the $\kappa$-mechanism}

   \author{T. Masseron\inst{1}\and K. Hawkins\inst{1,2}  }

   \institute{Institute of Astronomy, Madingley Road, Cambridge  CB3 0HA, UK
  \and Department of Astronomy, Columbia University, 550 W 120th St, New York, NY 10027\\
        \email{tpm40@ast.cam.ac.uk}}

   \date{Received ; accepted }

% \abstract{}{}{}{}{} 
% 5 {} token are mandatory
 
  \abstract
   {Stellar spectroscopy provides useful information on the physical properties of stars such as effective temperature, metallicity and surface gravity. However, those photospheric characteristics are often hampered by systematic uncertainties. The joint spectro-seismo project (APOGEE+Kepler, aka APOKASC) of field red giants has revealed a puzzling offset between the surface gravities ($\log g$) determined spectroscopically and those determined using asteroseismology, which is largely dependent on the stellar evolutionary status. }
   {Therefore, in this letter, we aim to shed light on the spectroscopic source of the offset.}
   {We used the APOKASC sample to analyse the dependencies of the $\log g$ discrepancy as a function of stellar mass and stellar evolutionary status. We discuss and study the impact of some neglected abundances on spectral analysis of red giants, such as He and carbon isotopic ratio. }
   {We first show that,  for stars at the bottom of
the red giant branch where the first dredge-up had occurred, the discrepancy between spectroscopic $\log g$ and asteroseismic $\log g$ depends on stellar mass. This seems to indicate that the $\log g$ discrepancy is related to CN cycling. Among the CN-cycled elements, we demonstrate that the carbon isotopic ratio ($\rm ^{12}C/^{13}C$) has the largest impact on the stellar spectrum. In parallel, we observe that this $\log g$ discrepancy shows a similar trend as the $\rm ^{12}C/^{13}C$ ratios as expected by stellar evolution theory. Although we did not detect a direct spectroscopic signature of $\rm ^{13}C$, other corroborating evidences suggest that the discrepancy in $\log g$ is tightly correlated to the production of $\rm ^{13}C$ in red giants. Moreover, by running the data-driven algorithm (the Cannon) on a synthetic grid trained on the APOGEE data, we try to evaluate more quantitatively the impact of various $\rm ^{12}C/^{13}C$ ratios.  }
   {While we have demonstrated that $\rm ^{13}C$ indeed impacts all parameters, the size of the impact is smaller than the observed offset in $\log g$. If further tests confirm that $\rm ^{13}C$ is not the main element responsible of the $\log g$ problem, the number of spectroscopic effects remaining to be investigated is now relatively limited (if any).}

   \keywords{
               }

   \maketitle
%
%-------------------------------------------------------------------

\section{Introduction}
\begin{figure}
   \centering
   \includegraphics[angle=-90,width=9cm]{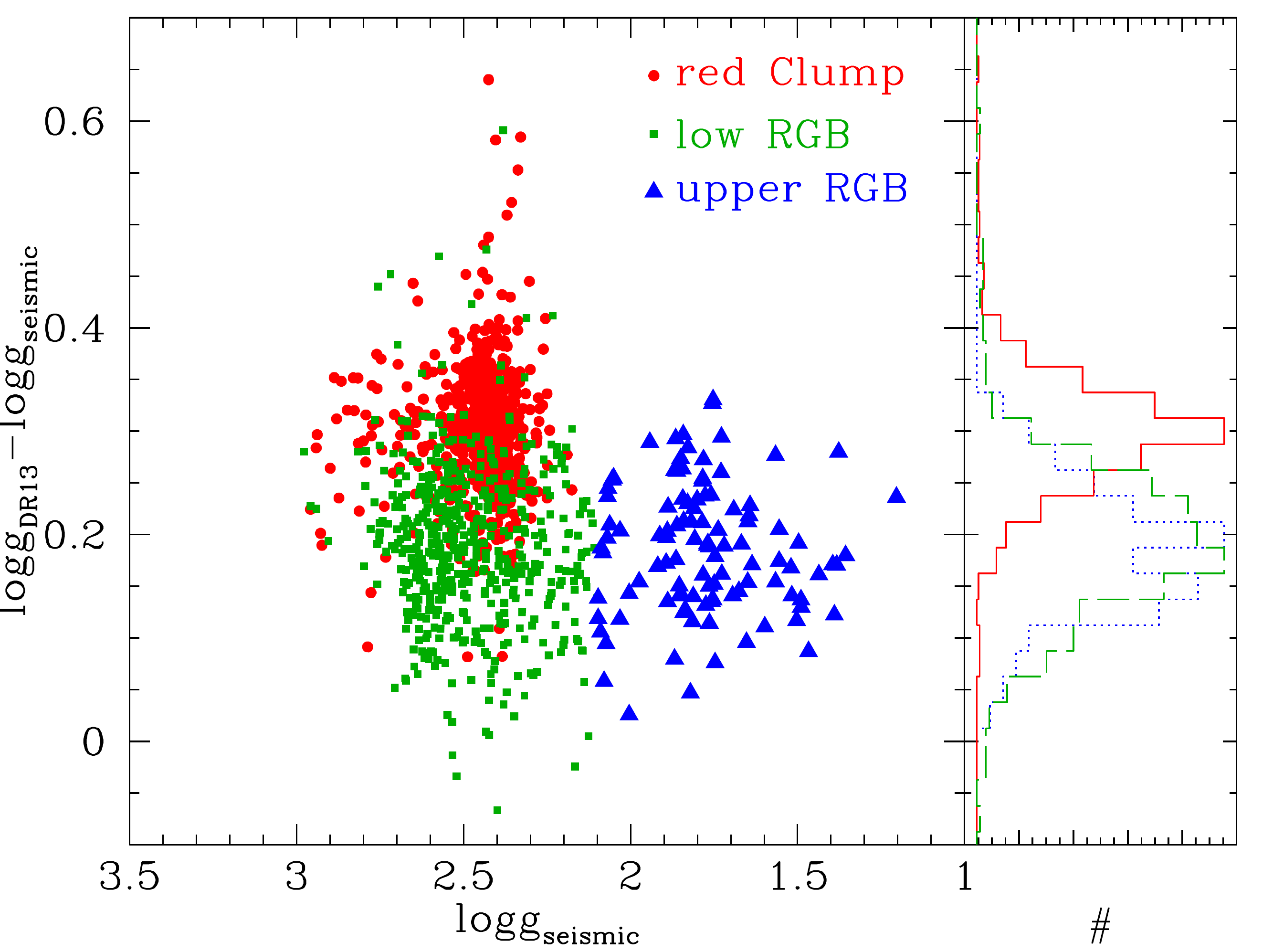}
     \caption{Difference in $\log g$ derived in the uncalibrated APOGEE DR13 $\rm\log$g and the $\rm\log$g obtained by asteroseismology. The sample is divided in three categories with three different symbols: RC stars as red points, lower RGB as green squares, and blue triangles as upper RGB.}
         \label{deltaloggvslogg_DR13}
   \end{figure}

In order to decipher stellar physics and Galactic populations, current stellar surveys such as Gaia-ESO \citep{Gilmore2012}, GALAH \citep{DeSilva2015}, LAMOST \citep{Luo2015} or SDSS/APOGEE \citep{SDSSDR132016} record several hundred thousand stellar spectra.  Due to the large number of observations, this naturally leads to the development of automatic procedures \citep[e.g.][]{Smiljanic2014,Ness2015,GarciaPerez2016,Masseron2016} to conduct their analysis. It was quickly realised that one of the most crucial aspect of those surveys resides in evaluating the performances of their internal procedures. The universally adopted method to evaluate the pipeline performances consists of comparing the results to the values of well-established reference stars. There are mainly three kinds of reference stars: the so-called benchmark stars \citep{Heiter2015}, stars in globular and open cluster and stars with asteroseismic data. A recent set of studies \citep[e.g. ][and reproduced in Fig.~\ref{deltaloggvslogg_DR13}]{Meszaros2013,Holtzman2015} have shown a puzzling feature when comparing the surface gravity ($\log g$) spectroscopic and the asteroseismic $\log g$ in the APOKASC dataset.  Indeed, this figure shows a bimodal discrepancy in surface gravity ($\log g$) obtained by spectroscopic and seismic methods depending on whether the star is at the core helium burning phase or at the H shell burning phase. However, it is not expected that the photosphere is influenced by ongoing internal nucleosynthesis. In other words, no distinct spectroscopic $\log g$ is expected between red clump (RC) stars and red giant branch (RGB) stars. Therefore, assuming this $\log g$ phenomenon is indeed a photospheric effect, understanding it represents an extraordinary opportunity to unambiguously distinguish the stellar evolutionary status of red giants from spectroscopy alone.  \\
In this letter, we attempt to narrow down the spectroscopic origin of this effect by constraining and testing unexplored spectroscopic variables such as abundances and isotopic ratios.    
%--------------------------------------------------------------------
\section{Data}
To carry out our analysis, we used the APOKASC sample \citep{Pinsonneault2014}. To build up the sample, we cross-matched the APOGEE DR13 uncalibrated stellar parameters from \citet{SDSSDR132016} with the asteroseismic masses and stellar parameters ($\rm T_{eff}, \log g$)  from \citet{Pinsonneault2014}, as well as the evolutionary status (Yvonne Elsworth, in prep.). The difference in the log g derived from spectroscopy and seismology as a function of asteroseismic log g for this sample is shown in Fig.~\ref{deltaloggvslogg_DR13}. We divide the sample into three categories according to asteroseismology: the RC stars, the low RGB stars, defined as RGB stars with $\rm logg >2.1$, and upper RGB stars, RGB stars with $\rm logg <2.1$. Note that, while we always display these groups, we do not discuss extensively this latter category of stars because those are significantly cooler stars and are possibly affected by other spectroscopic biases.
   \begin{figure}
   \centering
     \includegraphics[angle=-90,width=9cm]{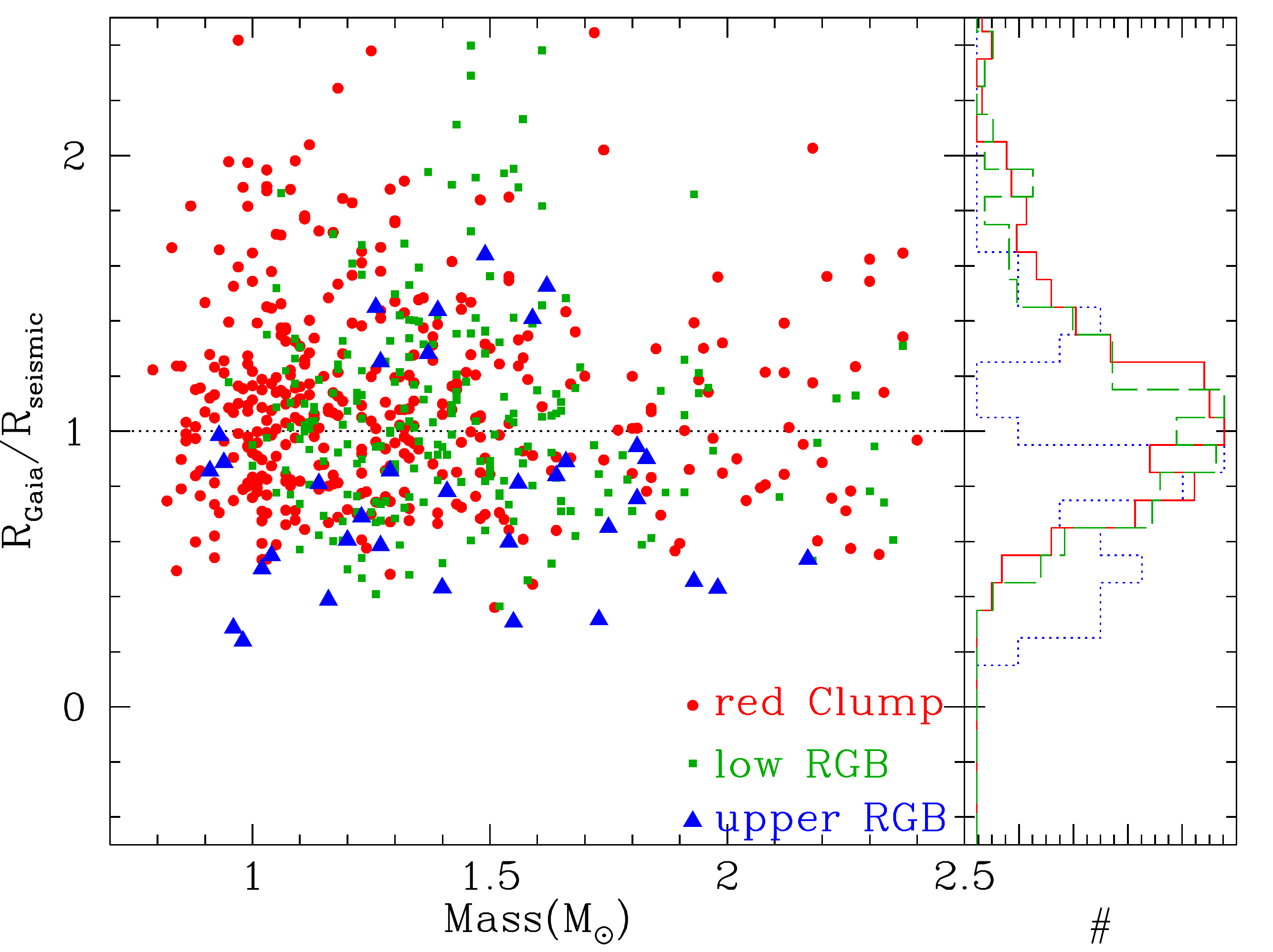}
    \caption{Comparison of radius obtained by asteroseismology and by parallax measurement for a subsample of our stars.}
         \label{GaiavsKepler_radius}
   \end{figure}

%-----------------------------------------------------------------
\section{Results and discussion}
First, we found evidence that the $\log g$ discrepancy is rather inherent to the spectral analysis. While an error in effective temperature could impact the seismic $\log g$ determination, the required change in $\rm T_{eff}$ to achieve a $\approx$0.2 offset in $\log g$ would be unrealistic \citep[$\sim$ 4000K by extrapolating from the study of ][]{Morel2014}. Therefore, another source of systematic error in the determination of the seismic $\log g$ has to be explored.  Thanks to the recent Gaia release \citep{Brown2016,Lindegren2016}, we have compared the radius values obtained using asteroseismology and those obtained using parallax. To compute the latter, we use the effective temperature as presented in the previous section as well as the bolometric corrections from \citet{Jordi2010}. In Fig.~\ref{GaiavsKepler_radius}, we compare the radius obtained via the two methods and note that both RGB and RC agree well. Therefore, we conclude that the asteroseismic analysis provides at least relatively consistent radii between RGB and RC stars. Thus, although the seismic mass may also be questioned \citep{Constantino2015,Gaulme2016}, we consider that the $\log g$ provided by asteroseismology is not driving the  $\log g$ discrepancy between RGB and RC stars.      \\
   \begin{figure}
   \centering
     \includegraphics[angle=-90,width=9cm]{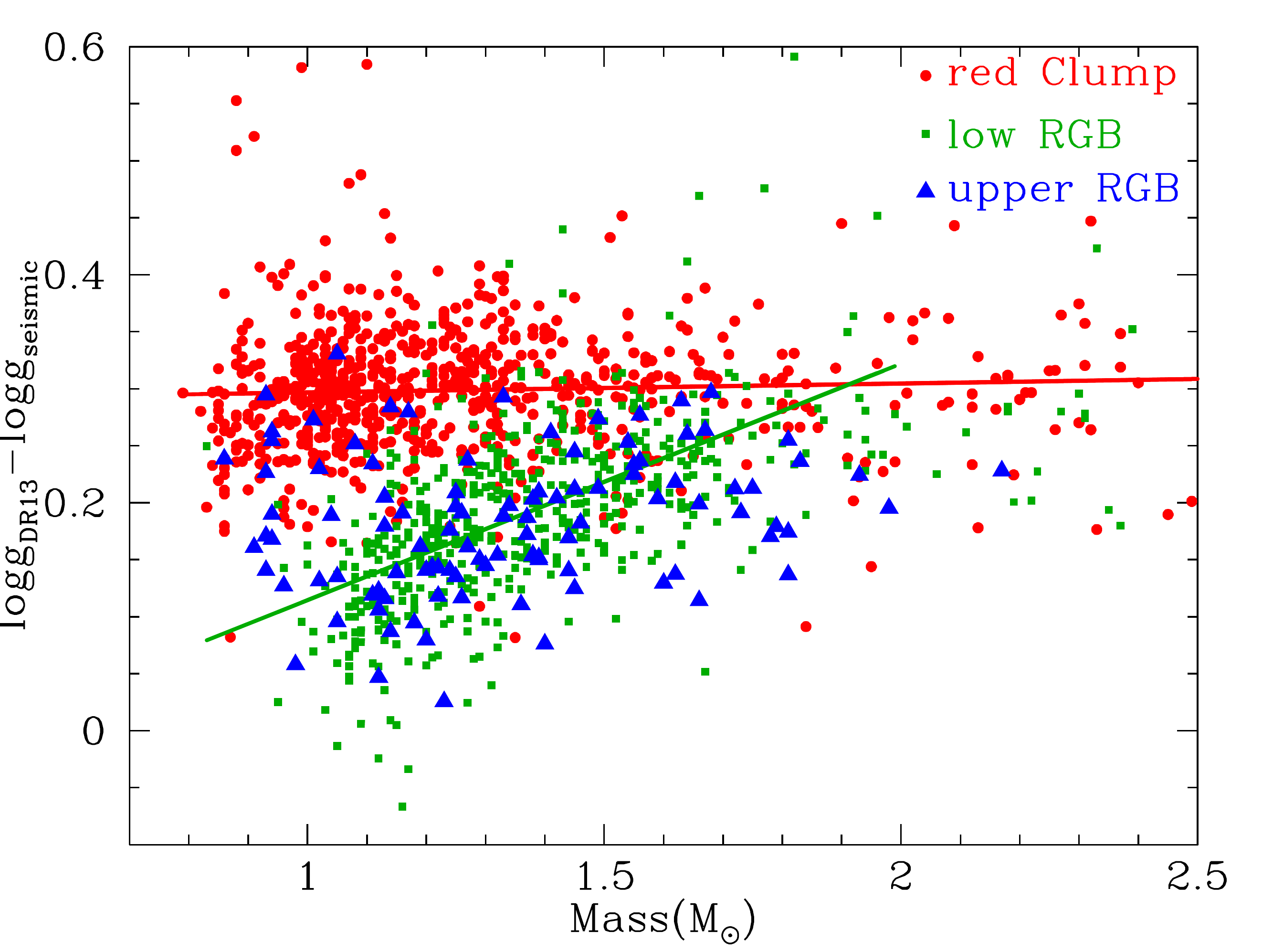}
    \caption{As in Fig.~\ref{deltaloggvslogg_DR13}, but as a function of stellar mass instead of $\log g$. The lines drawn are least square fit of the data points.}
         \label{deltaloggvsMass_DR13}
   \end{figure}
Holtzman et al. (in prep.) illustrated a metallicity dependency in the $\log g$ discrepancy. However, metallicity dependent parameter gradients in red giant stars may in fact underline a mass dependent gradient \citep{Masseron2016b}. Therefore we argue that it may be more relevant to study the mass dependency rather than metallicity. 
In Fig.~\ref{deltaloggvsMass_DR13}, we present the same discrepancy in $\log g$ as in Fig.~\ref{deltaloggvslogg_DR13} but as a function of stellar mass. While the largest discrepancy between the spectroscopic and seismic ($\Delta \log g$) are with the RCs, the most interesting new feature of this diagram is the correlation of $\Delta \log g$ with stellar mass for low RGB stars. The missing ingredient in the spectral analysis that leads to the discrepancy in $\log g$ between RGB and RC stars seems also tied to the only event that low RGB stars have undergone: the first dredge up. According to low-mass stellar evolution theory, the first dredge-up brings the product of H burning to the surface during the main sequence, and would thus affect H, He, Li, Be, B, C, N,  $\rm ^{12}C/^{13}C$, $\rm ^{14}N/^{15}N$, $\rm ^{16}O/^{17}O$, and  $\rm ^{16}O/^{18}O$ \citep{Iben1967,Kippenhahn1990,Charbonnel1994}. \\
   \begin{figure}
   \centering
   \includegraphics[angle=0,width=9cm]{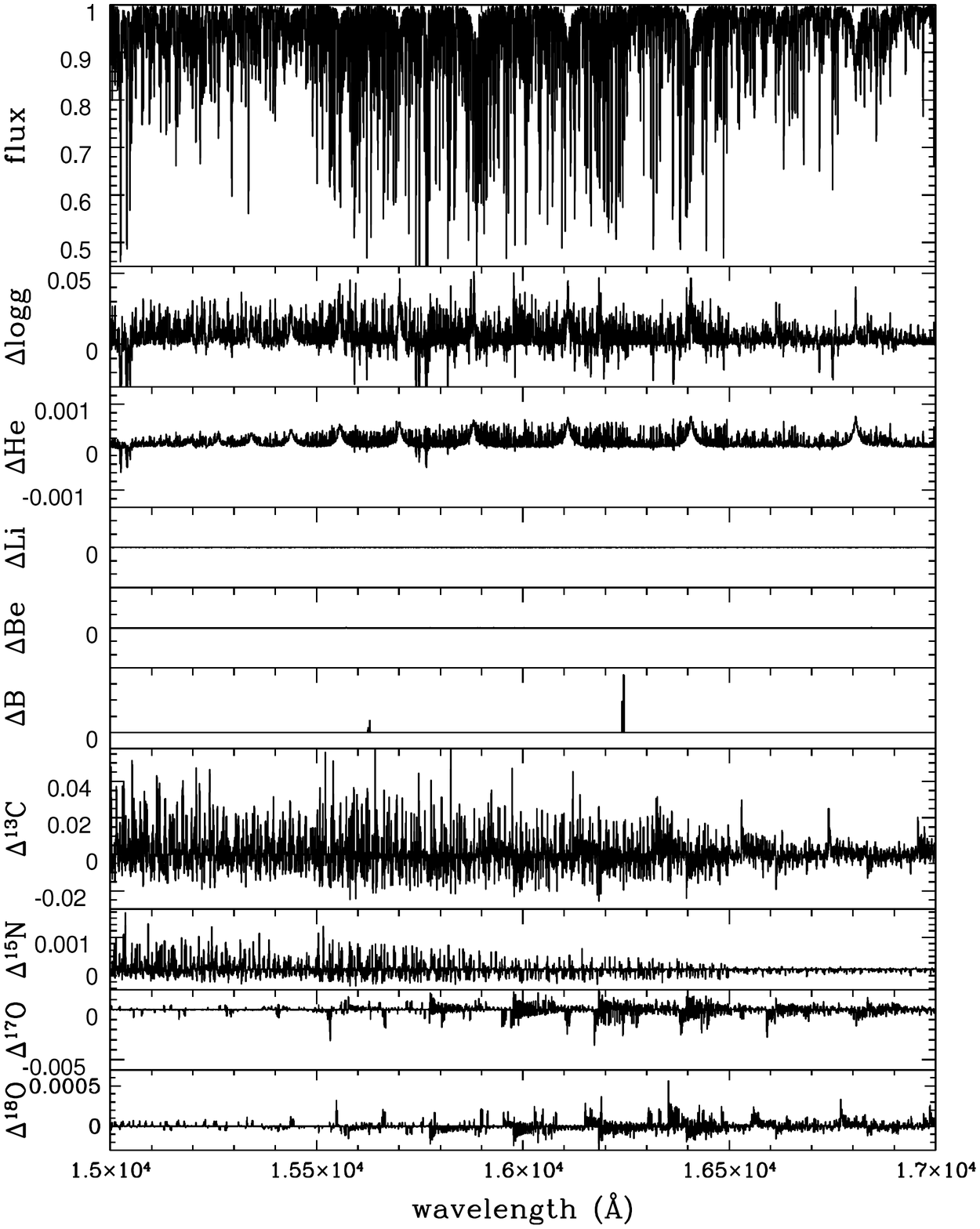}
      \caption{Normalised H-band spectrum for a star with $\rm  T_{eff}$=4500K, $\log g$=2.0, [C/H]=-0.3, [N/H]=+0.4 and solar composition model ([M/H]=0, [He/H]=0, [Li/H]=0, [Be/H]=0, [B/H]=0, $\rm^{12}C/^{13}C$=90, $\rm^{14}N/^{15}N$=272, $\rm^{16}O/^{17}O$=2625, $\rm^{16}O/^{18}O$=498) (top panel). The difference in normalised flux with the same parameters except, respectively, from upper to lower panel, $\rm \log g$=2.5, [He/H]=0.1, [Li/H]=-1.0, [Be/H]=-1.4, [B/H]=-2.7, $\rm^{12}C/^{13}C$=5, $\rm^{14}N/^{15}N$=100000, $\rm^{16}O/^{17}O$=60, and $\rm^{16}O/^{18}O$=25000.}              
         \label{4500g2.0z0.0_sample}
   \end{figure}
While C and N abundances are already taken into account in the APOGEE analysis \citep[with the ASPCAP pipeline,][]{GarciaPerez2016}, we show in Fig.~\ref{4500g2.0z0.0_sample} the impact on the spectrum of the other elements. Not surprisingly, Li and Be do not contribute to the spectrum and B shows only two very weak lines. The isotopic ratios of N and O contribute via the multiple CN, CO and OH molecular features over the wavelength range concerned, albeit relatively insignificantly given that their equilibrium values are very large \citep[$\rm^{14}N/^{15}N$=100000, $\rm^{16}O/^{17}O$=60, and $\rm^{16}O/^{18}O$=25000, ][]{Boothroyd1999}.  Therefore, the only remaining parameters that standard analysis did not take into account are He abundance and carbon isotopic ratio.\\
Helium appears to be a promising candidate as it impacts continuum opacities and mean molecular weight. Consequently, a change in He abundance alters the line profiles of H lines, line strength (and in particular molecular lines), ionisation equilibrium and molecular equilibrium \citep{BoehmVitense1979}, especially around 1.5$\mu$m where H$^-$ opacity reaches a minimum. However, the expected surface abundances along the red giant evolution are relatively small \citep[typically $\rm \lbrack$He/H$\rbrack$$\approx$+0.05-+0.1, ][corresponding to Y$\approx$0.27-0.30]{Lagarde2012}, yielding a relatively weak contribution compared to a change of $\log g$ of 0.5 dex.  We note, however, that in the particular case of He, one should also evaluate the change in the atmospheric structure. We do not, however, have the means to achieve that in this work. \\
 In contrast, a similar amplitude in the variation of the spectrum is obtained when changing the $\rm ^{12}C/^{13}C$ ratio from nearly solar values to typical RC values. In fact, because the total number of carbon atoms ($\rm N(^{12}C)+N(^{13}C)$) needs to be conserved in the spectral synthesis, the different spectrum is an addition of many weaker $\rm ^{12}C$ features and slightly enhanced $\rm^{13}C$ features.  However, those variations in flux do not systematically appear at the same wavelengths between a change in $\rm ^{12}C/^{13}C$  and a change in $\rm \log g$, and to precisely estimate its impact on stellar parameters using only the observation of the spectrum is not straightforward.\\
   \begin{figure}
   \centering
   \includegraphics[angle=-90,width=9cm]{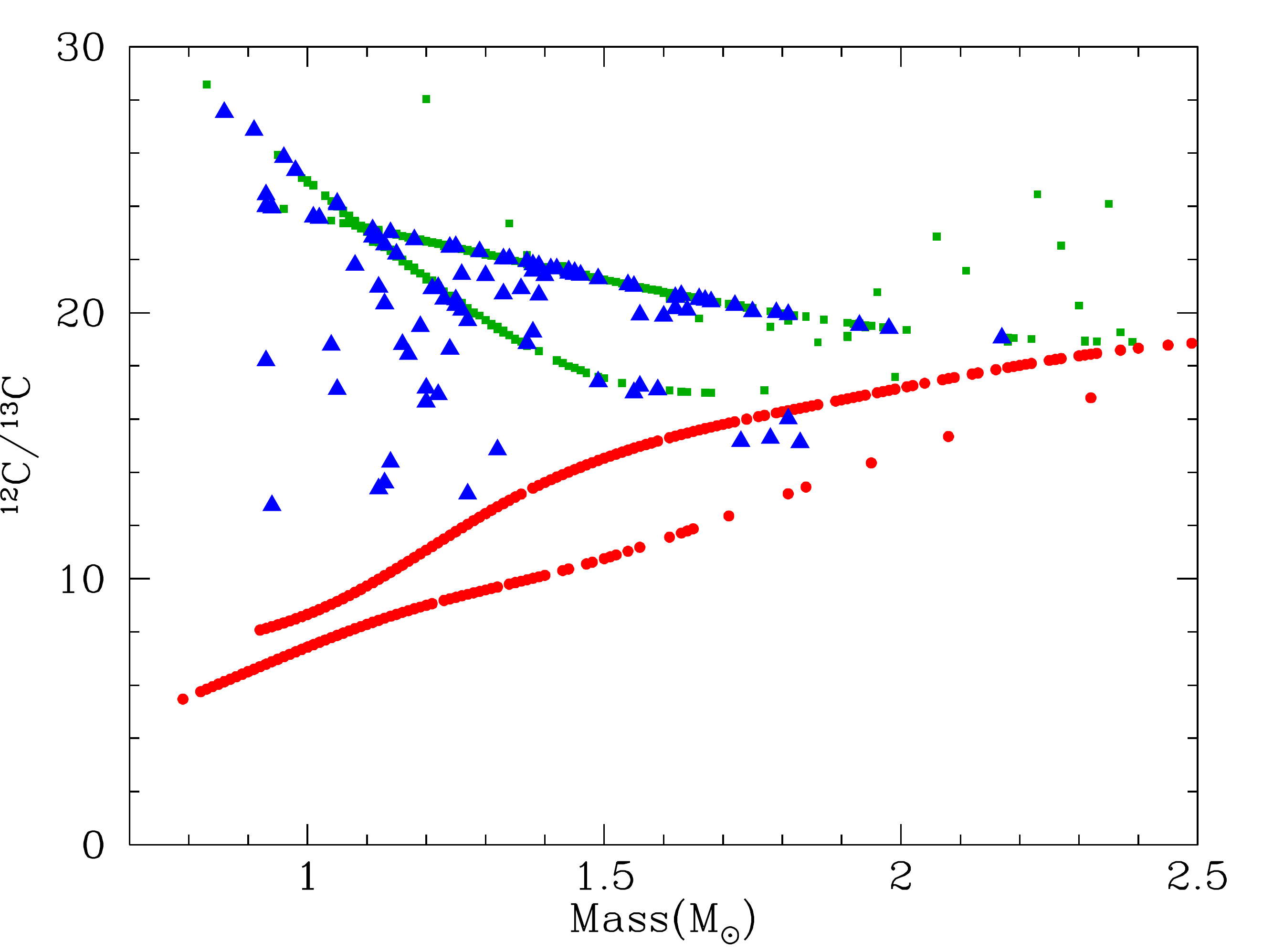}
    \caption{Theoretical surface $\rm ^{12}C/^{13}C$ ratio values for the APOKASC sample as a function of their asteroseismic mass derived from the stellar evolution models of \citet{Lagarde2012}. Those values are used as input for computation of the spectral grid.}
         \label{C12C13models}
   \end{figure}
 To more quantitatively evaluate the impact of the change of $\rm ^{12}C/^{13}C$ ratio on the stellar parameters, we built a grid of synthetic spectra matching the APOKASC sample using Turbospectrum \citep{Plez2012}. We adopted the same stellar parameters and atomic and molecular linelists as described in \citet{Hawkins2016}.  We computed the grid for two $\rm ^{12}C/^{13}C$ isotopic ratios: a generic value (15 as assumed for all giants in the ASPCAP grid) and a theoretical value for each star. This theoretical $\rm ^{12}C/^{13}C$ value is derived by interpolating its observed mass and $\rm T_{eff}$ within the theoretical grid of \citet{Lagarde2012}. The derived theoretical values for the satrs in our sample are shown in Fig.~\ref{C12C13models}. Note that, because the metallicity of the models are too coarse, we adopted the closest models to the star's metallicity. Knowing that the main sample covers a metallicity range of $\rm-0.8~<~[Fe/H]~<~0.5$, we chose the Z=0.004 models for stars with $\rm [Fe/H]<-0.2$ and the Z=0.014 models for the stars with $\rm [Fe/H]>-0.2$.  This results in a bimodal trend for low RGB and RC stars in Fig.~\ref{C12C13models} that are nonetheless in the same direction.  We then ran the Cannon \citep[a pipeline that employs a data-driven approach to determine stellar parameters and chemical abundances, ][]{Ness2015} on this synthetic grid to derive their stellar parameters using the uncalibrated DR13 stellar parameters and C, N and $\alpha$ element abundances for a sample of 5000 stars as our training set. This training set was employed using the same criteria as in \citet{Casey2016} with an additional criterion that the uncalibrated parameters and C and N abundances must also be known. Once the cannon was trained, the stellar parameters (T$\rm_{eff}, \log$g, [M/H], $\rm[\alpha/Fe]$, [C/H], and [N/H]) of the spectra could be inferred. \\
   \begin{figure}
   \centering
   \includegraphics[angle=0,width=9cm]{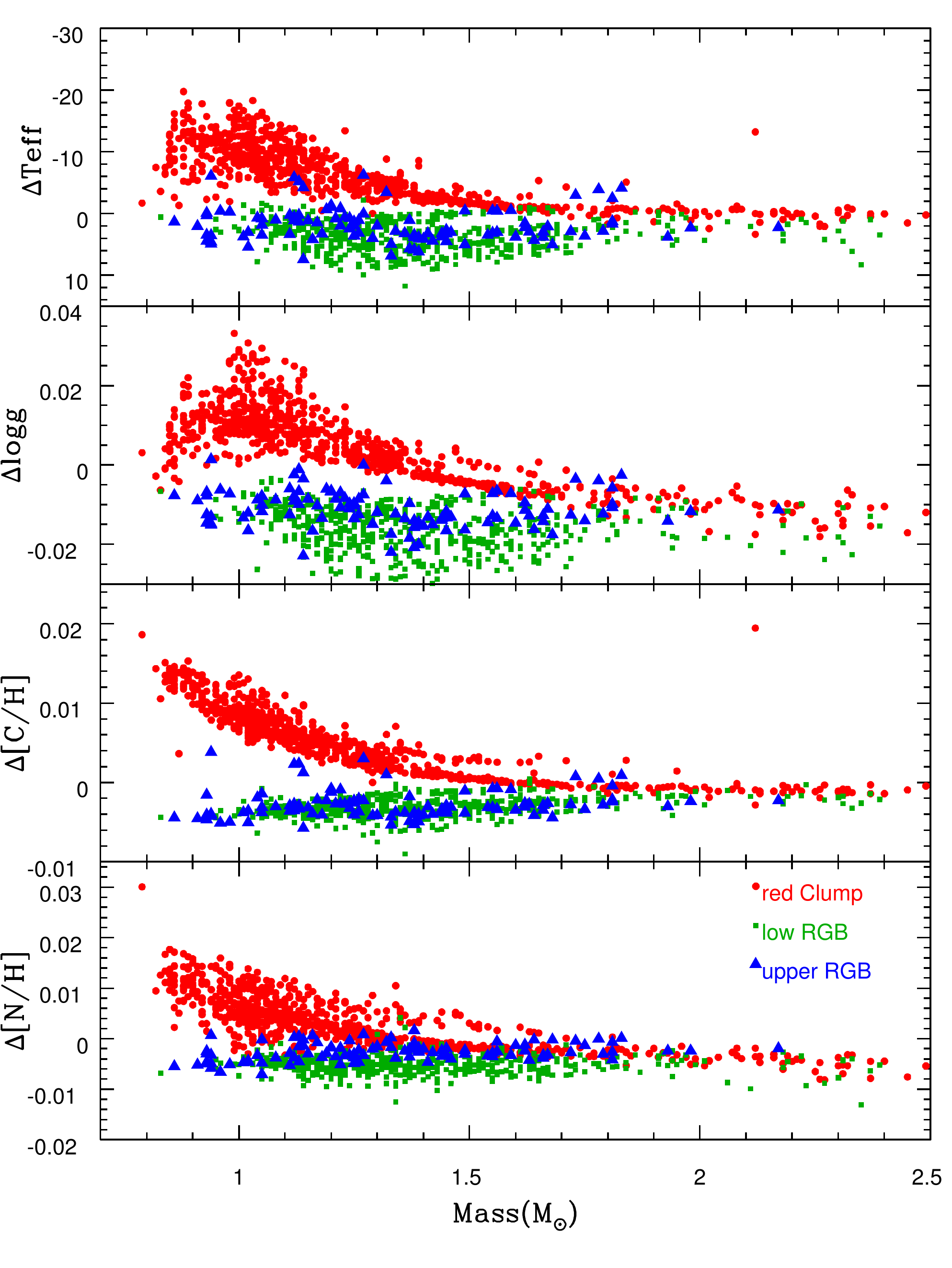}
      \caption{ Resulting difference in stellar parameter determination ($\rm T_{eff}, \log g$, [C/H], and [N/H]) obtained by the Cannon on synthetic spectra when assuming a $\rm ^{12}C/^{13}C$ ratio of 15 and the stellar evolution model values described in \citet{Lagarde2012}, and illustrated in Fig.~\ref{C12C13models}. }
         \label{deltaTeffvsMass_Cannon}
   \end{figure}
Fig.~\ref{deltaTeffvsMass_Cannon} shows the difference in the stellar parameters obtained by the Cannon when changing the $\rm ^{12}C/^{13}C$ ratio. First, it is interesting in this figure to point out the similarities with Fig.~\ref{deltaloggvsMass_DR13}. In particular, the RC and RGB stars are easily distinguishable. This is expected as they have distinct $\rm ^{12}C/^{13}C$ ratios (Fig.~\ref{C12C13models}). This distinct  $\rm ^{12}C/^{13}C$ ratio between the low RGBs and the RCs is explained by the occurrence of non-canonical extra mixing between the two stellar evolutionary stages. During this extra mixing along the upper RGB branch, CN-cycled material is pulled up to the surface decreasing the $\rm ^{12}C/^{13}C$ ratio. It is worth noticing that the extra mixing is more effective at lower masses.  However, the overall change between RGB and RC in $\log g$ due to the change in $\rm ^{12}C/^{13}C$ ratio is approximately 0.05 dex, which is approximately four time larger in the observed APOGEE spectra (Fig.~\ref{deltaloggvsMass_DR13}). Therefore, we conclude that the missing spectroscopic factor that could explain the $\log g$ problem is probably not $\rm^{13}C$. We also demonstrate earlier that it is even less likely to be He, Li, Be, B, nitrogen or oxygen isotopic ratios.

In any case, we observe that a change in $\rm ^{12}C/^{13}C$ ratio impacts not only $\log g$ but all stellar parameters.  This may indicate that while $\rm^{13}C$ does contribute to the discrepancy in $\rm \Delta\log g,$ it is probably not the only parameter. We stress here that our procedure is not the exact pipeline employed by the APOGEE team and thus we would suggest the collaboration considers the isotopic ratio. Nonetheless, our results demonstrate that including $\rm ^{12}C/^{13}C $  ratio in the spectral analysis of red giants is recommended for an improvement in the determination of stellar parameters. Furthermore, when considering, notably, the N abundance, this may lead to systematic effects that can be as large as the quoted precision. 
\section{Conclusions}
We show that the discrepancy between the $\log g$ derived spectroscopically in the APOGEE data and the $\log g$ derived by asteroseismology seems related to CN cycling in red giants. If this problem is due to a missing parameter in the spectral analysis, it is very likely to be $\rm ^{13}C,$ but when simulating the $\rm^{13}C$ effect on stellar parameters, the obtained shift in $\Delta \log g$ is not as large as the observed one. Naturally, this may be a result of the fact that we use a different procedure than the ASPCAP pipeline. One first, obvious, required future test is to run the APOGEE pipeline with a grid that includes the $\rm ^{12}C/^{13}C$  ratio as an extra variable. Another direction is to measure $\rm ^{12}C/^{13}C$ ratio in a sample of those giants. However, it is difficult to find strong enough lines to measure in the APOGEE spectra. Counterpart optical spectra are thus required.\\
 We believe that, in this letter, we have explored several spectroscopic effects that may explain the presence of an offset in $\log g$. Nevertheless, if the present results are confirmed, this may question assumptions made concerning the yields from stellar evolution and the $\log g$ from asteroseismology. 

\begin{acknowledgements}
This work was partly supported by the European Union FP7 programme through
ERC grant number 320360. KH thanks the Marshall Scholarship program and the Simons Society of Fellows
\end{acknowledgements}

%\begin{thebibliography}
\bibliographystyle{aa} % style aa.bst
\bibliography{RGBRC} % your references RGBRC.bib
%\end{thebibliography}
%
%
\end{document}